\documentclass[twocolumn,pra,aps,showpacs]{revtex4}
\usepackage{graphicx}
\begin{document}

\title{Cross-dimensional relaxation in Bose-Fermi mixtures}

\author{J. Goldwin}
\author{S. Inouye}
\altaffiliation{Present address: Department of Physics, University
of California, Berkeley, CA 94720, USA.}
\author{M. L. Olsen}
\author{D. S. Jin}
\altaffiliation{Quantum Physics Division, National Institute of
Standards and Technology}
\affiliation{JILA, National Institute of
Standards and Technology, and Department of Physics, University of
Colorado, Boulder, Colorado 80309}
\date{\today}

\begin{abstract}
We consider the equilibration rate for fermions in Bose-Fermi
mixtures undergoing cross-dimensional rethermalization. Classical
Monte Carlo simulations of the relaxation process are performed
over a wide range of parameters, focusing on the effects of the
mass difference between species and the degree of initial
departure from equilibrium. A simple analysis based on Enskog's
equation is developed and shown to be accurate over a variety of
different parameter regimes. This allows predictions for mixtures
of commonly used alkali atoms.
\end{abstract}

 \maketitle
\maketitle

\section{Introduction\label{sec:intro}}

In the study of ultracold dilute atomic gases, the interparticle
interactions are characterized by the $s$-wave scattering length
$a$. Because of the low temperatures achieved, and the lack of
long range or anisotropic interactions, the value of $a$
determines a wide variety of equilibrium and dynamical properties
of quantum degenerate gases. Furthermore the parameter $a$ can be
tuned in many systems, from $-\infty$ to $\infty$ by means of
Feshbach resonances \cite{FRtheory}, allowing the experimenter
access to any desired interaction strength.  This potential for
real-time control over interactions is a unique feature of
ultracold gas experiments. Additionally the efficiency of
evaporative cooling relies on a large elastic collision rate,
which is proportional to $a^2$. For these reasons, the ability to
accurately determine the scattering properties of dilute ultracold
gases becomes essential for quantum gas experiments.

In their pioneering work with ultracold $^{133}$Cs atoms, Monroe
and co-workers showed that a relatively simple rethermalization
measurement starting with a non-equilibrium gas could provide a
determination of the elastic collision cross-section~\cite{Mon93},
which is equal to $8\pi a^2$ for identical non-condensed bosons.
The rethermalization rate for a gas in the so-called collisionless
regime (defined by a collision rate per particle much lower than
the harmonic frequencies of the trapping potential), was measured
by selectively removing energy from the gas in one spatial
dimension and watching the subsequent cross-dimensional
rethermalization. The relaxation in such experiments is driven by
elastic collisions, and an analysis based on Enskog's equation
shows that the rate of relaxation is proportional to the mean rate
of collisions \cite{Rob01,Rei65},
\begin{equation}
\Gamma_{\rm relax} = \frac{1}{\alpha}\,\langle n\rangle
\,\sigma\,\langle v_{\rm rel}\,\rangle \quad , \label{eq:GammaB}
\end{equation}
where $n$ is the number density of the gas, $\sigma$ is the
elastic collision cross-section, and $v_{\rm rel}$ is the relative
collision speed; brackets $\langle\cdot\rangle$ denote a thermal
average, and we have used the assumptions of energy-independent
$s$-wave collisions and Boltzmann statistics to write the mean
collision rate $\langle\,\Gamma_{\rm coll}\,\rangle = \langle n\,
\sigma \,v_{\rm rel}\rangle=\langle n\rangle\, \sigma\, \langle
v_{\rm rel}\rangle$. The constant of proportionality $\alpha$,
defined as the ratio of collision and relaxation rates, reflects
the mean number of collisions per particle required for
rethermalization. A variety of numerical and analytical studies
have found that $\alpha$ is between $2.5$ and $2.7$
\cite{Mon93,Rob01,Wu96,Kav98,DeM99}.

We have recently extended the method of cross-dimensional
relaxation to probe the $s$-wave scattering length $a_{BF}$
between bosonic and fermionic atoms in a Bose-Fermi mixture
\cite{Gol04}. In that work, we compared single-species boson
rethermalization to relaxation of fermions in the presence of
bosons in order to eliminate the systematic calibration error
associated with atom number that typically dominates
cross-dimensional rethermalization measurements. Since collisions
between spin-polarized ultracold fermions are forbidden by the
Pauli exclusion principle \cite{DeM99}, their rethermalization
proceeds only through collisions with the bosons. The collision
rate per fermion in the mixture therefore depends only on the
number of bosons, allowing us to write the mean relaxation rate
per fermion in analogy with the single-species case,
\begin{equation}
\Gamma_F = \frac{1}{\beta}\,\langle n_B \rangle\, \sigma_{BF}
\,\langle v_{BF}\rangle \quad . \label{eq:GammaF}
\end{equation}
Here $\langle n_B \rangle$ is the equilibrium density of the
bosons, averaged over the fermion distribution, $\sigma_{BF}$ is
the boson-fermion elastic cross section, and \mbox{$\langle
v_{BF}\rangle=(8\,k_B T/\pi \mu)^{1/2}$} is the thermally averaged
relative collision speed between bosons and fermions with reduced
mass \mbox{$\mu = m_F\,m_B/(m_F+m_B)$} and temperature $T$ ($k_B$
is Boltzmann's constant). The cross-section is related to the
interspecies scattering length by $\sigma_{BF}=4\pi a_{BF}^2$. The
constant of proportionality $\beta$ reflects the mean number of
collisions per fermion needed for rethermalization. Knowledge of
$\alpha$ and $\beta$ was essential in achieving both the accuracy
and precision of the measurement in Ref. \cite{Gol04}.

In this work, we study the general dependence of $\beta$ on the
masses of the fermions and bosons by means of detailed classical
Monte Carlo simulations of the relaxation process. We additionally
address the effect due to the finite initial departure from
equilibrium. Understanding this effect is a necessary component of
the analysis of both the simulations and the measurements, where a
larger initial perturbation improves the signal-to-noise ratio. We
further develop a classical kinetic model based on Enskog's
equation for the case of Bose-Fermi mixtures that reproduces the
behavior of the simulations.  The analysis and simulations show
the following: (i) Equation (\ref{eq:GammaF}) is valid for a wide
range of parameters relevant to current experiments, (ii) the
difference in mass between the bosons and fermions can lead to a
$\sim 5$ times difference in $\beta$ between light and heavy
fermions in mixtures of experimental interest, and (iii) the size
of the initial perturbation must be taken into account to fully
understand the results.

The remainder of the paper is organized as follows. In Sec.
\ref{sec:montecarlo} we describe the details of the Monte Carlo
simulations and use the simulations to verify the validity of
Eq.(\ref{eq:GammaF}). In Sec. \ref{sec:model} the mass dependence
of the relaxation is investigated using the simulations and
compared to a classical kinetic theory that is developed in some
detail. It is shown that the model reproduces the behavior of the
simulations over a wide range of masses. In Sec.
\ref{sec:discussion} the importance of the size of the initial
perturbation is addressed. Finally, we conclude in Sec.
\ref{sec:conclusion} with a discussion of possible extensions of
the model and further applications of the Monte Carlo simulations.

\section{Monte Carlo Simulation\label{sec:montecarlo}}

Implicit in the definition of $\beta$ in Eq.(\ref{eq:GammaF}) are
the assumptions that the energy anisotropy undergoes exponential
decay \cite{Wu96} and that $\beta$ depends only on intrinsic
properties of the particles under study, such as the mass and
quantum statistics, and not the bulk properties of the gases, such
as the temperature or atom numbers (assuming one remains always in
the collisionless regime). These assumptions were verified by
running Monte Carlo simulations of the relaxation.

In the experiment considered here, we start with a Bose-Fermi
mixture that is in thermal equilibrium in a cylindrically
symmetric harmonic trap and increase the radial trapping frequency
by a factor of $\Omega$ to produce the initial energy anisotropy
in the system; the axial frequency is unchanged. The change in
trap strength is performed slowly with respect to the radial trap
periods but quickly compared to the thermal relaxation time.  The
increase in radial energies $E_{x,y}$ is then the same for each
species and is equal to the fractional increase in the trapping
frequencies. Therefore the ratio of radial and axial energies
immediately before rethermalization is simply
\begin{eqnarray*}
\left.\frac{E_{x,y}}{E_z}\right|_{t=0} =\Omega \quad .
\end{eqnarray*}
Note that for gases prepared in this manner, $E_x\approx E_y$ at
all times. Since the gases rethermalize together, they reach the
same final temperature. In the classical gas limit, the final
temperature is given by $(T_\infty/T_0) = (1+2\,\Omega)/3$, where
$T_0$ is the temperature before compression. It is important to
note that for this type of rethermalization, there is no net
transfer of energy between species during the relaxation process;
this fact is essential for the validity of Eq.(\ref{eq:GammaF}).

An outline of the Monte Carlo simulation is as follows. An
ensemble consisting of $N_B$ bosons and $N_F$ fermions is prepared
in a cylindrically symmetric harmonic trapping potential. The
ensemble is initialized by assigning a random position and
velocity vector to each particle from separable Gaussian
distributions. The distributions are scaled such that there is a
factor of $\Omega$ imbalance between the mean energies per
particle (both kinetic and potential) in the transverse ($x,y$)
and axial ($z$) dimensions. The initial energy in a given
dimension, however, is the same for both species. The positions
and velocities are evolved for some small time step $\Delta t$
according to Newton's laws in the trap, and then collisions are
considered.

Only boson-boson and boson-fermion collisions are allowed, in
accordance with the Pauli exclusion principle for a system with
spin-polarized fermions. If two particles are found within a
critical distance $r_c$ of each other, the pair is given a chance
to collide. The collision probability is given by $P_{\rm coll} =
\sigma_{kl}\,v_{\rm rel}\, \Delta t/{\mathcal V}_c$, where
$\sigma_{kl}$ is the collision cross section between particles $k$
and $l$, $v_{\rm rel} = |\vec{v}_k-\vec{v}_l|$ is the relative
collision speed, and ${\mathcal V}_c = 4\pi\, r_c^3/3$ is the
volume of the sphere containing the colliding atoms. If $P_{\rm
coll}$ is greater than a uniformly distributed random number
between zero and one, then the relative velocity vector is rotated
into a random direction (conserving the total momentum and energy)
to represent an $s$-wave collision. Finally, after all possible
collision pairs have been considered, the mean energies of each
species in each of the three cartesian directions are recorded,
and a new time step proceeds.

In agreement with Ref.\cite{Gue99} we find it essential for
obtaining robust results to set $\Delta t$ and $r_c$ such that the
ensemble average collision probability $\langle P_{\rm
coll}\rangle$ and occupation of the fictional spheres surrounding
the atoms are both well below $10\%$. Typical values for our
simulations are $\Delta t\sim 10^{-3}$ to $10^{-2}$ times the
collision time $\langle n\,\sigma\,v_{\rm rel}\rangle^{-1}$, and
$r_c \sim 10^{-3}$ to $10^{-2}$ times the root-mean-squared (rms)
cloud radius. Since the motion between collisions is known
analytically, it is not necessary to keep $\Delta t$ small
compared to the trap period. Finally, we note that we have not
observed any double-counting of collisions in consecutive time
steps. This is because our time step $\Delta t$ typically
satisfies $v\,\Delta t\gg r_c$, where $v$ is the characteristic
speed of the particles.

\begin{figure}
\includegraphics[scale=0.8]{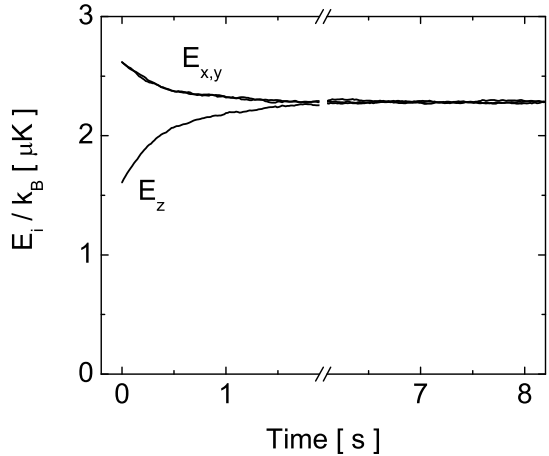}
\caption{Typical Monte Carlo simulation of cross-dimensional
relaxation of fermions in the presence of bosons. The mean energy
per fermion $E_i$ in each direction is shown versus time during
the equilibration process. The curves represent $5 \times 10^4$
fermions in the presence of $5 \times 10^4$ bosons. The
calculation used $^{40}$K ($^{87}$Rb) as the fermion (boson), and
the trapping frequencies, temperature, and initial perturbation
were taken from the experiment in Ref. \cite{Gol04}. A
cross-species scattering length of $|a_{BF}|=235\,a_0$, with $a_0$
the Bohr radius, was assumed. The calculated mean time between
collisions is \mbox{$0.23$ s}.\label{fig:decay}}
\end{figure}

A typical relaxation curve obtained from the simulation is shown
in Fig. \ref{fig:decay}. The relaxation rate $\Gamma_F$ of the
energy anisotropy of the fermions is determined by fitting the
ratio of radial and axial energies to the ratio of decaying
exponentials,
\begin{eqnarray*}
\frac{1}{2}\,\frac{E_x+E_y}{E_z}= \frac{1+\varepsilon
\exp(-\Gamma_F \,t)}{1-2\,\varepsilon \exp(-\Gamma_F\,t)} \quad .
\end{eqnarray*}
Here, $E_i$ is the mean total energy per particle of the fermions
in the $i^{\rm th}$ direction, and $\varepsilon=(\Omega
-1)/(1+2\Omega)$ characterizes the amount of initial perturbation
from equilibrium.

\begin{figure}\begin{center}
\includegraphics[scale=0.8]{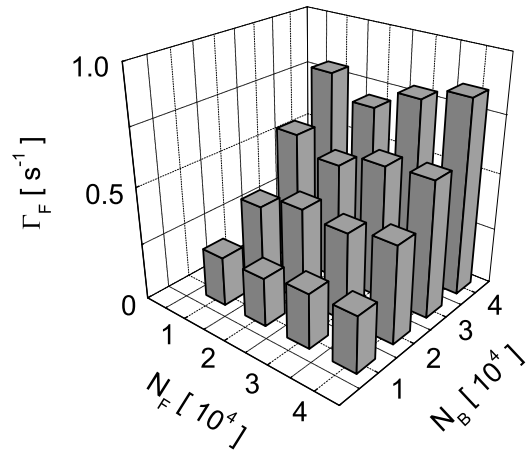}
\caption{Fermion rethermalization rate $\Gamma_F$ as a function of
$N_B$ and $N_F$. These calculations used the parameters from the
$^{87}$Rb-$^{40}$K experiment in Ref.\cite{Gol04}, with
$|a_{BF}|=235\,a_0$. \label{fig:barchart}}\end{center}
\end{figure}

The validity of Eq.(\ref{eq:GammaF}) was tested by running Monte
Carlo simulations with varying temperatures, numbers of atoms,
trapping frequencies, and interspecies cross-sections
$\sigma_{BF}$. As an example, the dependence of $\Gamma_F$ on the
number of bosons ($N_B$) and fermions ($N_F$) in the mixture is
shown in Fig. \ref{fig:barchart}. The results show that $\Gamma_F$
varies linearly with $N_B$, while it is constant when changing
$N_F$ over the same range.  Although these simulations assumed
$|a_{BF}|=235\,a_0$ and $a_{BB} = 98.98\,a_0$ \cite{Kem02},
results similar to those in Fig. \ref{fig:barchart} were obtained
for $a_{BF}=50\,a_0$, corresponding to a ratio of cross-sections
$(\sigma_{BB}/\sigma_{BF})=0.35$ and $7.8$, respectively.

\section{Mass Dependence --- Classical Kinetic Model\label{sec:model}}

The Monte Carlo simulation is a powerful tool for analyzing the
behavior of relaxing Bose-Fermi mixtures, but it requires a
considerable amount of computing resources. For example, one of
our simulations using $10^4$ fermions and $10^4$ bosons with
$10^4$ time steps took over 8 hours on a Unix system with a $2.4$
GHz processor and 4 GB of RAM. The computation time, which is
dominated by the pairwise search for collision partners, scales
roughly as the product of the number of time steps and
$N_B\,(N_F+N_B/2)$, so that simulations become impractical for
$N\gtrsim 10^5$. Furthermore, it is desirable to build up a
physical picture of the relaxation that is not provided by the
simulations.  For these reasons, we now consider an analytic model
of the rethermalization.

\begin{figure}
\includegraphics[scale=0.8]{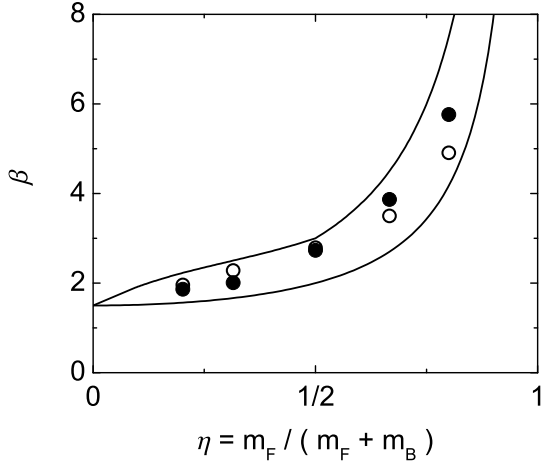}
\caption{Dependence of the mean number of collisions per fermion
needed for rethermalization on the normalized fermion mass. Solid
points ($\bullet$) are the results of the Monte Carlo simulations,
and open points ($\circ$) are from exponential fits to our model
from Eq.(\ref{eq:fullresult}).  The solid lines are the bounding
predictions from the classical kinetic model. The number of atoms
is $N_B=N_F=10^4$, the initial perturbation is $\Omega=1.64$, and
the ratio of cross sections is $(\sigma_{BB}/\sigma_{BF})=0.35$.
\label{fig:massdep}}
\end{figure}

One new degree of freedom for two-species rethermalization is the
appearance of a second mass.  In order to see how the number of
collisions per fermion needed for rethermalization depends on the
masses of the particles involved, we first consider the two
limiting cases of very light and very heavy fermions. Recall that
energy is redistributed by means of $s$-wave collisions, which
randomize the relative velocities of the colliding particles.  For
very light fermions, a single collision should be nearly enough to
redistribute the energy. If the fermions are far heavier than the
bosons, however, it should take many collision times to
redistribute the energy, since a single collision has little
effect on the motion of the heavy particle. We therefore expect
the number of collisions needed for equilibration to be a
monotonically increasing function of the normalized fermion mass
$\eta=m_F/(m_F+m_B)$. Note that this is in contrast to the
behavior of two gases initially at different (isotropic)
temperatures and subsequently brought into contact for
rethermalization.  In the latter case, one expects
$\beta^{-1}\propto 4\,\eta\,(1-\eta)$ \cite{Mos01}, which is most
efficient for $\eta=1/2$. The results of our Monte Carlo
simulations is shown as the solid points in Fig.
\ref{fig:massdep}. As expected, $\beta$ increases smoothly with
the fermion mass. The open points and solid lines are predictions
from a simple kinetic model, which we now discuss in detail.

Our analysis of the rethermalization process is based on Enskog's
equation, which is equivalent to the Boltzmann transport equation
\cite{Rei65}. This treatment gives the rate of change of the
ensemble average of any function of the boson and fermion
positions and velocities, usually denoted
$\chi(\vec{x}_B,\vec{v}_B;\vec{x}_F,\vec{v}_F)$, by
\begin{eqnarray}
\dot{\langle\,\chi\,\rangle} = \sigma\,\langle\,n\,v_{\rm
rel}\,\Delta\chi\,\rangle \quad , \label{eq:enskog}
\end{eqnarray}
where we have again used the assumption of energy-independent
$s$-wave scattering to separate the collision cross-section
$\sigma$ from the ensemble average. The quantity $\Delta\chi$ is
just the change in $\chi$ due to a single collision. Here we
consider
\begin{eqnarray}
\chi_1 \equiv E_{1x}-E_{1z} \quad , \label{eq:chidefinition}
\end{eqnarray}
where we are focusing for now on generic particles of type 1
colliding with particles of type 2. The energy $E_{1i}$ denotes
the total (kinetic plus potential) energy of type-1 atoms in the
$i^{\rm th}$ direction. Note that our current interest in $\chi_1$
implies the ensemble average in Eq.(\ref{eq:enskog}) is taken only
over the distribution function for type-1 atoms.  Based on the
results of Fig. \ref{fig:barchart}, we can assume equal numbers of
atoms $N_1=N_2$ without loss of generality.

Immediately after a collision, only the kinetic energy (KE) has
changed, so that
\begin{eqnarray*}
\Delta\chi_1 &=& \Delta ({\rm KE}_{1x} - {\rm KE}_{1z}) \\
&=& \frac{1}{2}\,m_1\,\Delta (v_{1x}^2-v_{1z}^2) \quad .
\end{eqnarray*}
Since $\Delta\chi_1$ has no position dependence, we can remove the
type-2 particle density $n_2$ from the average, yielding
\begin{eqnarray*}
\dot{\langle\,\chi_1\,\rangle} = \frac{1}{2}\,m_1\, \langle\,
n_2\,\rangle\,\sigma_{12}\, \langle\,v_{\rm rel}\,\Delta
(v_{1x}^2-v_{1z}^2)\, \rangle \quad .
\end{eqnarray*}

We define CM and relative velocities in the usual manner,
\begin{eqnarray}
\nonumber \vec{V}_{\rm CM} &=& \frac{m_1\vec{v}_1 + m_2\,\vec{v}_2}{m_1+m_2} \\
\vec{v}_{\rm rel} &=& \vec{v}_1 - \vec{v}_2 \quad ,
\label{eq:cmandrel}
\end{eqnarray}
which gives
\begin{eqnarray*}
v_{1x}^2-v_{1z}^2 &=& \left( V_{{\rm CM}x}^2 - V_{{\rm CM}z}^2
\right) \\
&\quad& + \left( \frac{m_2}{m_1+m_2} \right)^2 \left(v_{{\rm
rel}x}^2 - v_{{\rm rel}z}^2 \right) \\
&\quad& +\,2\,\frac{m_2}{m_1+m_2}\left(V_{{\rm CM}x}\,v_{{\rm
rel}x} - V_{{\rm CM}z}\,v_{{\rm rel}z} \right) \quad .
\end{eqnarray*}
Since the collision leaves $\vec{V}_{\rm CM}$ and $|\vec{v}_{\rm
rel}|$ unchanged, but randomly rotates the direction of
$\vec{v}_{\rm rel}$, we are left with
\begin{eqnarray*}
\dot{\langle\,\chi_1\, \rangle} &=& -
\frac{1}{2}\, \frac{m_1\,m_2}{(m_1+m_2)}\,\langle\, n_2\,\rangle\,\sigma_{12} \\
&\quad& \times\,\left\langle v_{\rm
rel}\left[\,\frac{m_2}{m_1+m_2}\,\left(v_{{\rm rel}x}^2
- v_{{\rm rel}z}^2 \right) \right.\right. \\
&\quad& \left.\left. + \frac{}{}\,2\left( V_{{\rm CM}x}\,v_{{\rm
rel}x} - V_{{\rm CM}z}\,v_{{\rm rel}z}
\right)\,\right]\,\right\rangle \quad .
\end{eqnarray*}
Here we need only consider the quantities immediately before the
collision since there is no preferred direction for $\vec{v}_{\rm
rel}$ after the collision.  Note also that $\langle\, V_{{\rm
CM}i}\,v_{{\rm rel}i}\, \rangle \ne 0$ for a gas with a
cross-dimensional energy anisotropy.

Calculating these ensemble averages for arbitrary masses and
energy anisotropies (which are in general different between
species) is non-trivial, but some simple approximations may be
used. For the first term it is easy to show that for Boltzmann
distributions in equilibrium one has
\begin{eqnarray}
\langle\, v_{\rm rel}\,v_{{\rm rel}x}^2 \, \rangle = \frac{4}{3}\,
\langle\,v_{\rm rel}\,\rangle\, \langle\, v_{{\rm rel}x}^2\,
\rangle \quad . \label{eq:vrelvrelx2}
\end{eqnarray}
For cross-terms of the form $\langle\, V_{{\rm CM}x}\,v_{{\rm
rel}x}\,\rangle$, which vanish under equilibrium conditions, one
can still consider the limit of small deviation from thermal
equilibrium. In this case we obtain, similar to the above result,
\begin{eqnarray*}
\langle\, v_{\rm rel}\,V_{{\rm CM}x}\,v_{{\rm rel}x}\, \rangle &=&
\frac{4}{3}\,\left\langle\,v_{\rm rel}\,\rangle\,\langle\,V_{{\rm
CM}x} \, v_{{\rm rel}x}\,\right\rangle \quad .
\end{eqnarray*}
We verified by analytic means and with Monte Carlo integrations of
Gaussian distributions that these approximations are reasonable
for small anisotropies.

Combining these results gives
\begin{eqnarray*}
\dot{\langle\,\chi_1\, \rangle} &=& -\frac{2}{3}\,\langle\,
\Gamma_{12}\,\rangle\, \frac{m_1\,m_2}{m_1+m_2}\,
\left\langle\,\frac{m_2}{m_1+m_2}\,\left(v_{{\rm rel}x}^2
- v_{{\rm rel}z}^2 \right) \right. \\
&\quad& \left. + \frac{}{}\,2\left( V_{{\rm CM}x}\,v_{{\rm rel}x}
- V_{{\rm CM}z}\,v_{{\rm rel}z} \right)\,\right\rangle \quad .
\end{eqnarray*}
The collision rate $\Gamma_{12}$ describes the rate per type-1
particle of collisions with particles of type 2. If we finally
substitute back with $v_1$ and $v_2$, and use $\langle\,v_{1i}\,
v_{2i}\,\rangle = 0$, we obtain
\begin{eqnarray}
\dot{\chi}_1 = -\frac{2}{3}\,\Gamma_{12}\,\frac{m_2}{(m_1+m_2)^2}
\, \left[\,(2\,m_1 + m_2)\,\chi_1 - m_1\,\chi_2\,\right]
\label{eq:chi1}
\end{eqnarray}
where we have dropped the angle brackets now for simplicity.  Note
that we have used the fact that the mean kinetic and potential
energies in a given direction are equal in the collisionless
regime.

For concreteness, we now associate the type-1 particles with the
fermions and type-2 particles with the bosons.  The
time-dependence of $\chi_2$, defined in analogy with
Eq.(\ref{eq:chidefinition}), is obtained by swapping
$1\leftrightarrow 2$ in Eq.(\ref{eq:chi1}) and adding a term
describing the effect of boson-boson collisions. This term was
calculated from Enskog's equation in Ref.\cite{Rob01}, giving
$\alpha = 5/2$ in the limit of small energy anisotropy.  This
yields the final result,
\begin{eqnarray} \nonumber
\frac{d}{d\,\tau}\;\chi_1 &=&
-\frac{2}{3}\,\frac{m_2}{(m_1+m_2)^2}\,
\left[\,(2\,m_1 + m_2)\,\chi_1 - m_1\,\chi_2\,\right] \\
\nonumber \frac{d}{d\,\tau}\;\chi_2 &=&
-\frac{2}{3}\,\frac{m_1}{(m_1+m_2)^2} \, \left[\,(2\,m_2 +
m_1)\,\chi_2 -
m_2\,\chi_1\,\right] \label{eq:fullresult} \\
&\quad& -\frac{\gamma}{\alpha}\,\chi_2 \quad ,
\end{eqnarray}
where we have introduced the dimensionless time
\mbox{$\tau=\Gamma_{12}\,t$} and the ratio of collision rates
$\gamma = \Gamma_{22}/\Gamma_{12}$ (note that
$\Gamma_{12}=\Gamma_{21}$ for Boltzmann distributions with
$N_1=N_2$ under conditions of thermal equilibrium).

Equation (\ref{eq:fullresult}) is the main result of our analysis.
The results of exponential fits to the time-evolution of
Eq.(\ref{eq:fullresult}) are shown as the open points in Fig.
\ref{fig:massdep}.  The slight overestimation (underestimation)
for light (heavy) fermions is a result of our approximation in
Eq.(\ref{eq:vrelvrelx2}).  Although the full solution to
Eq.(\ref{eq:fullresult}) requires prior knowledge of the
inter-species cross-section $\sigma_{12}$ (through $\gamma$), we
now show that the solutions in the limits $\gamma=0$ and
$\gamma\to\infty$ provide tight bounds on $\beta$. These two
limits are shown as the solid lines in Fig. \ref{fig:massdep}.

In the case $\gamma \gg 1$, where boson-boson collisions occur
much more frequently than boson-fermion collisions, we can assume
the energy of the bosons is always isotropic and take $\chi_2=0$
for all times. In this limit, Eq.(\ref{eq:fullresult}) is easily
solved, giving
\begin{eqnarray*}
\chi_1(\tau) = \chi_1(0) \exp\left(-\frac{\tau}{\beta_l}\right)
\quad.
\end{eqnarray*}
with $\beta_l$ given by
\begin{eqnarray}
\beta_l = \frac{3}{2}\,\frac{1}{1-\eta^2} \quad , \label{eq:betaL}
\end{eqnarray}
The solution to Eq.(\ref{eq:fullresult}) in the limit $\gamma\to
0$ does not correspond exactly to a simple exponential but rather
to the sum (or difference) of decaying exponentials.  Heavy
fermions ($\eta > 1/2$) rethermalize slowly at first and then more
quickly as the energy of the bosons becomes isotropic.  In this
case, we obtain an upper limit on $\beta$ by simply taking the
small-$\tau$ expansion of the solution to
Eq.(\ref{eq:fullresult}), giving $\beta_u=(3/2)(1-\eta)^{-1}$. For
light fermions ($\eta < 1/2$), we obtain $\beta_u$ by fitting the
solution of Eq.(\ref{eq:fullresult}) to the simple exponential
decay minimizing the chi-squared error. Our values for $\beta_u$
and $\beta_l$ are shown in Fig. \ref{fig:massdep} as the upper and
lower solid lines, respectively. The results from the Monte Carlo
simulations fall within these limits as expected.  Simulations
with different atom numbers and cross-sections gave similar
results.

\section{Discussion\label{sec:discussion}}

Before proceeding, we first consider the effect of the finite size
of the initial perturbation. Our analysis has so far been limited
to the case of arbitrarily small initial anisotropies.  As
discussed in the introduction, however, experiments (as well as
Monte Carlo simulations) require large anisotropies for good
signal-to-noise ratios. It was noted in Ref.\cite{Wu96} that the
mean collision rate per particle $\langle\,\Gamma_{\rm
coll}\,\rangle =\langle n\rangle\,\sigma\,\langle v_{\rm rel}
\rangle$ actually changes during the relaxation process, because
of the redistribution of energy. In Eq.(\ref{eq:GammaF}) we have
defined the relaxation rate in terms of the final equilibrium
collision rate. Because of the exponential nature of the
relaxation, however, the bulk of rethermalization occurs at
initial times. We therefore expect that the observed number of
collisions required for equilibration will be approximately
rescaled by a factor of $\langle\,\Gamma_{\rm coll} (t\to\infty)
\,\rangle / \langle\, \Gamma_{\rm coll} (t=0)\,\rangle$. As shown
in Appendix \ref{app:lambda}, this effect can be accounted for by
multiplying $\beta$ by
\begin{eqnarray}\nonumber
\lambda(\Omega) &=& \Omega\,
\left(\frac{3}{1+2\,\Omega}\right)^{3/2}\sqrt{\frac{4}{3}\,(1+2\,\Omega)} \\
 &\quad& \times \left[1+\Omega\,\frac{\tan^{-1} \sqrt{\Omega
-1}}{\sqrt{\Omega-1}}\; \right]^{-1} . \label{eq:lambda}
\end{eqnarray}
The function $\lambda (\Omega)$ goes to zero at either limit of
$\Omega$ \mbox{($0$ or $\infty$)} and reaches a smooth maximum of
$1$ at \mbox{$\Omega=1$}.

\begin{figure}
\includegraphics[scale=0.75]{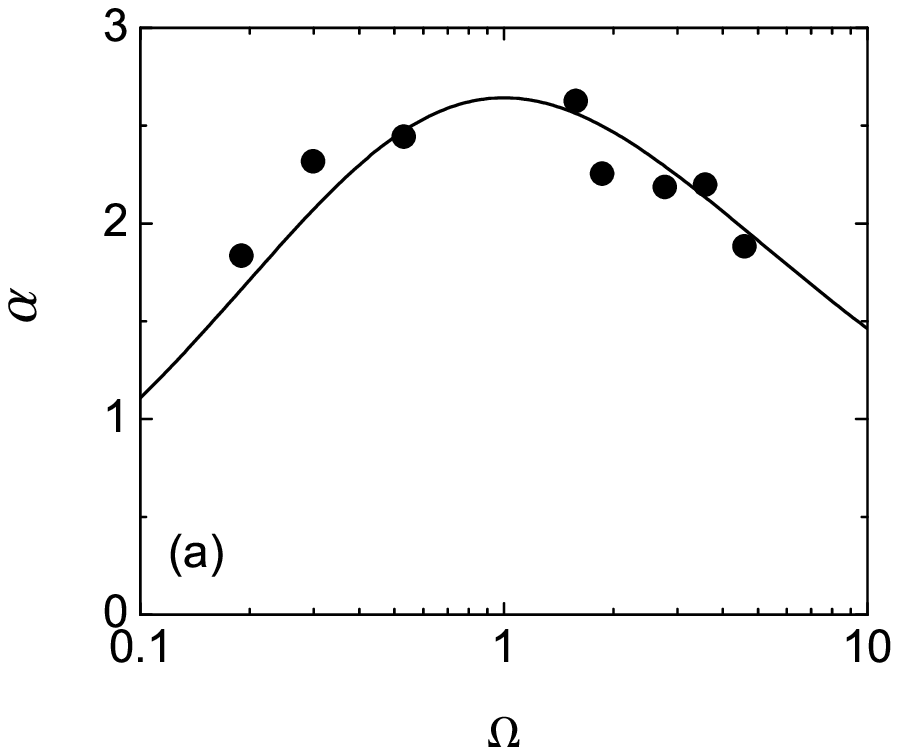}\hfill
\includegraphics[scale=0.75]{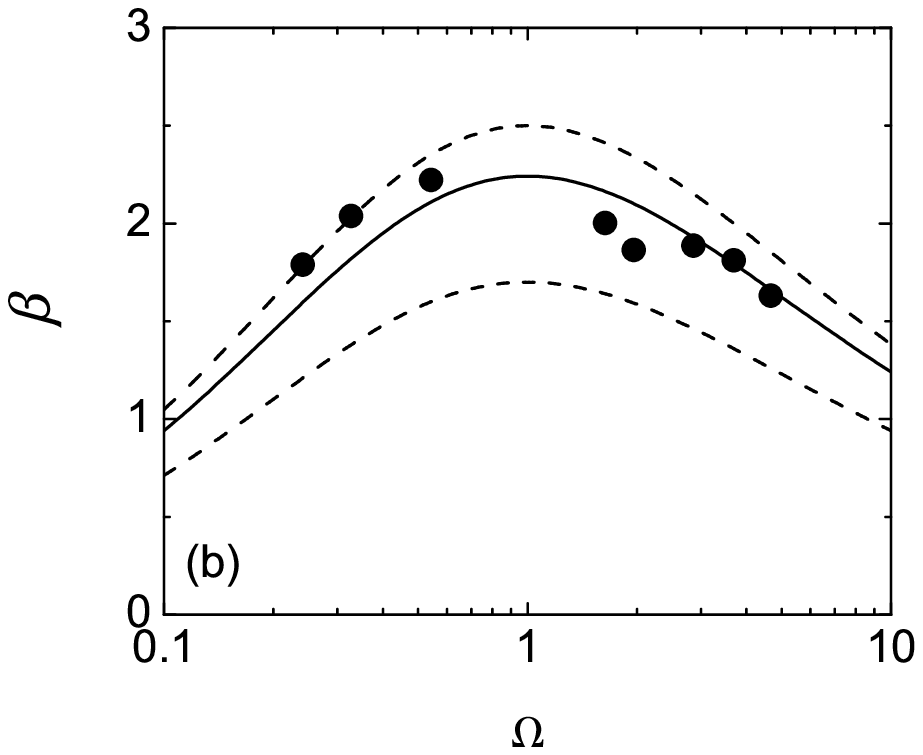}
\caption{Ratio of rethermalization and collision rates as a
function of the initial perturbation. (a) Single-species
relaxation of $^{87}$Rb. The points are results from the Monte
Carlo simulations, and the solid line is the best fit to
Eq.(\ref{eq:lambda}), which gives $\alpha = 2.64 \pm 0.07$ for the
limit of small initial anisotropy. (b) Number of collisions per
fermion ($^{40}$K) for rethermalization with $^{87}$Rb.  Points
are from the Monte Carlo simulations, and the upper and lower
dashed lines combine Eq.(\ref{eq:lambda}) with $\beta_u$ and
$\beta_l$, respectively. The solid line is the best fit to
Eq.(\ref{eq:lambda}), giving $\beta=2.26\pm 0.09$ for vanishing
initial anisotropy. \label{fig:anisotropy}}
\end{figure}

To test our prediction for $\lambda (\Omega)$ given by
Eq.(\ref{eq:lambda}), we performed a series of Monte Carlo
simulations, both for single-species relaxation of $^{87}$Rb and
for relaxation of $^{40}$K atoms in the presence of $^{87}$Rb, as
shown in Fig. \ref{fig:anisotropy}. The model and the Monte Carlo
show excellent agreement over a wide range of $\Omega$.

Finally, we predict in Table \ref{tab:predict} values of $\beta$
for various mixtures of bosonic and fermionic alkali atoms. The
nominal value is the average of $\beta_u$ and $\beta_l$ and the
uncertainty is half the difference. To date, all Bose-Fermi
mixtures produced in experiments have light fermions ($\eta <
1/2$), where the maximum uncertainty in our prediction is about
$20\%$. Since measurements of the magnitude of the scattering
length depend on $\beta^{-1/2}$ \cite{Gol04}, the model will
introduce only a small uncertainty in any determination of
$|a_{BF}|$ as described in this work.  As a test of the
performance of the prediction, we consider the $^{40}$K-$^{87}$Rb
system, for which our model predicts $\beta = 2.1\pm 0.4$. Monte
Carlo simulations over a wide range of $N_B,N_F$, and $\Omega$
gave $\beta=2.23\pm 0.07$ (in the limit $\Omega=1$), in excellent
agreement with the prediction.

\begin{table}
\begin{tabular}{|c||c|c|}\hline

~ & $^6$Li & $^{40}$K \\ \hline\hline

$^7$Li & $2.4\pm 0.5$ & $\;8\pm 2\;$ \\ \hline

$^{23}$Na & $1.9\pm 0.3$ & $3.3\pm 0.8$ \\ \hline

$^{41}$K & $1.8\pm 0.2$ & $2.5\pm 0.5$ \\ \hline

$^{87}$Rb & $1.64\pm 0.14$ & $2.1\pm 0.4$ \\ \hline

$^{133}$Cs & $\;1.60\pm 0.09\;$ & $\;1.9\pm 0.4\;$ \\ \hline
\end{tabular}
\caption{Predicted values of $\beta$ for Bose-Fermi mixtures of
commonly used alkali atoms. The nominal value is the average of
$\beta_u$ and $\beta_l$, and the uncertainty is equal to half of
the difference $\beta_u-\beta_l$.} \label{tab:predict}
\end{table}

\section{Conclusion\label{sec:conclusion}}
We have investigated the cross-dimensional rethermalization of
fermions in the presence of bosons, focusing attention on the
effects of the mass difference between species and the finite
departure from equilibrium necessary for experiments. Monte Carlo
simulations were performed over a wide range of parameters, and a
simple analysis based on Enskog's equation was developed that
reproduces the results from the simulations. We have further used
the model to predict the number of collisions per fermion needed
for rethermalization for a variety of alkali Bose-Fermi mixtures.

In this work, we have restricted ourselves to rethermalization due
to energy-independent $s$-wave collisions, but the Monte Carlo
simulations can easily be extended to include $p$-wave collisions
\cite{DeM99}, resonant scattering \cite{Arn97}, or damping of more
complicated collective excitations \cite{Gue99}. In addition, the
simulations have allowed us to investigate various effects not
accounted for in our model, such as the differential gravitational
sag for species with different masses \cite{note:sag}, and the
possibility of slightly different initial energies and
anisotropies between species.  The simulations could be trivially
extended to Fermi-Fermi \cite{DeM99} or Bose-Bose mixtures
\cite{Blo01} and could accommodate in a straightforward way
relaxation in the hydrodynamic regime, Bose-Einstein and
Fermi-Dirac distributions \cite{Lop97}, heating and loss during
the rethermalization, and the effects of anharmonic trapping
potentials.

\appendix\section{Calculation of $\lambda(\Omega)$\label{app:lambda}}
As discussed in Sec. \ref{sec:discussion}, the rescaling
$\lambda(\Omega)$ is given by the ratio of the final and initial
collision rates,
\begin{eqnarray*}
\lambda (\Omega) = \frac{\langle n_B(t\to\infty)\rangle}{\langle
n_B(t=0)\rangle} \, \frac{\langle v_{\rm
rel}(t\to\infty)\rangle}{\langle v_{\rm rel} (t=0)\rangle}\quad .
\end{eqnarray*}
We assume here that the mean energy in a given direction is the
same for fermions and bosons at the beginning and end of the
relaxation, and denote these energies
\begin{eqnarray}
\frac{(E_x,E_y,E_z)}{k_B\,T_\infty} = \left\{ \begin{array}{cl}
  \displaystyle \frac{3}{1+2\,\Omega}\;(\Omega,\Omega,1) & ,\quad t=0 \\
  ~&~\\
  (1,1,1) & ,\quad t\to\infty
\end{array} \right.
\label{eq:initconds}
\end{eqnarray}

A simple calculation for Gaussian distributions gives the mean
boson density (averaged over the fermion distribution function),
\begin{eqnarray*}
\langle n_B \rangle &\propto& (E_x\,E_y\,E_z)^{-1/2}\quad ,
\end{eqnarray*}
which, using Eq.(\ref{eq:initconds}), gives
\begin{eqnarray}
\frac{\langle n_B(\infty)\rangle}{\langle n_B(0)\rangle} =
\Omega\left(\frac{3}{1+2\,\Omega}\right)^{3/2} \quad .
\label{eq:density}
\end{eqnarray}
For $\langle v_{\rm rel} \rangle$ one finds for our initial
conditions,
\begin{eqnarray*}
\langle v_{\rm rel}(0) \rangle &=&
\sqrt{\frac{3}{1+2\,\Omega}\,\frac{2\,k_BT_\infty}{\pi\,\mu}}\;\left[\,1
+
\frac{\Omega\,\tan^{-1}\sqrt{\Omega-1}}{\sqrt{\Omega-1}}\;\right]
\end{eqnarray*}
where $\mu=m_F\,m_B/(m_F+m_B)$ is the reduced mass, as in the
text. Comparing to the equilibrium value gives
\begin{eqnarray}
\frac{\langle v_{\rm rel}(\infty)\rangle}{\langle v_{\rm rel
}(0)\rangle} = \sqrt{\frac{4}{3}\,(1+2\,\Omega)}
\left[1+\Omega\,\frac{\tan^{-1} \sqrt{\Omega
-1}}{\sqrt{\Omega-1}}\; \right]^{-1} \label{eq:fullv}
\end{eqnarray}
Note that for $\Omega < 1$ the fraction in the square brackets can
be written $(1-\Omega)^{-1/2}\tanh^{-1}(1-\Omega)^{1/2}$, which is
again real-valued. The product of Eqs.(\ref{eq:density}) and
(\ref{eq:fullv}) was presented as $\lambda(\Omega)$ in
Eq.(\ref{eq:lambda}).

\begin{acknowledgments}
We gratefully acknowledge useful discussions with C. Wieman, E.
Cornell, and P. Engels. This work was funded by a grant from the
U. S. Department of Energy, Office of Basic Energy Sciences, and
the National Science Foundation. The computing cluster used for
these simulations was provided by a grant from the W. M. Keck
Foundation.
\end{acknowledgments}


\begin{thebibliography}{99}
\bibitem{note:sag}
As mentioned in Ref. \cite{Gol04}, the mean collision rate per
fermion is reduced by a temperature-dependent factor describing
the different equilibrium positions of the gases in the trap
because of gravity.  We have verified with the Monte Carlo
simulations that the observed value of $\beta$ is unaffected by
gravity under realistic conditions when this reduction in overlap
between the gases is included in the collision rate.
\end{thebibliography}
\end{document}